\documentclass[aps,twocolumn,superscriptaddress,showpacs]{revtex4}
 \usepackage{amsmath}
 \usepackage{amssymb}
 \usepackage{amsfonts}
 \usepackage{graphicx}
 \usepackage{multirow}
 \usepackage{bm}

 \usepackage{dcolumn}
 \usepackage{bm}
 \usepackage{graphicx}
 \usepackage{color}
 \DeclareGraphicsExtensions{. jpg,. pdf, . mps, . png, . eps, . ps, . EPS}

\begin{document}
 \def \be{ \begin{equation}}
 \def \ee{ \end{equation}}
 \def \bc{ \begin{center}}
 \def \ec{ \end{center}}
 \def \bea{ \begin{eqnarray}}
 \def \eea{ \end{eqnarray}}
 \newcommand{ \avg}[1]{ \langle{#1} \rangle}
 \newcommand{ \Avg}[1]{ \left \langle{#1} \right \rangle}
\newcommand{\tc}[1]{\multicolumn{1}{c|}{#1}} 
\newcommand{\tl}[1]{\multicolumn{1}{l}{#1}} 
\newcommand{\tr}[1]{\multicolumn{1}{r}{#1}}

 \title{Entropy rate of non-equilibrium  growing networks}

 \author{Kun Zhao}
 \affiliation{Department of Physics, Northeastern University, Boston 02115 MA, USA}
 \author{Arda Halu}
 \affiliation{Department of Physics, Northeastern University, Boston 02115 MA, USA}
 \author{Simone  Severini  }
 \affiliation{Department of Computer Science, and Department of Physics  \& Astronomy,University College London, WC1E 6BT London, UK}
 \author{ Ginestra Bianconi}
 \affiliation{Department of Physics, Northeastern University, Boston 02115 MA, USA}
 \begin{abstract}
New entropy measures have been recently introduced for the quantification of the complexity of networks. Most of these entropy measures apply to static networks or to dynamical processes defined on static complex networks. In this paper we define   the entropy rate of growing network models. This entropy rate quantifies how many labeled networks are typically generated by the growing network models. We analytically evaluate the difference between  the entropy rate of growing tree network models and  the entropy of tree networks that have the same asymptotic degree distribution. We find that   the growing networks with linear preferential attachment  generated by dynamical models are exponentially less than the static networks with the same degree distribution for a large variety of relevant growing network models. We study the entropy rate for growing network models showing structural phase transitions including models with non-linear preferential attachment. Finally, we bring numerical  evidence that the entropy rate above and below the structural phase transitions follow a different scaling with the network size.
 \end{abstract}
 \maketitle
 \section{Introduction}
Complex networks describe the structure of many biological, social and technological systems  \cite{Albert2002,Dorogovtsev2003,Boccaletti2006,Dorogovtsev2010,Newman2010}. Recently, new entropy measures have been introduced for the quantification of complexity of networks  \cite{Sole2004,BGS,BGS1,Bianconi2008a,Bianconi2008b,Bianconi2009,Annibale2009,Anand2009,Anand2010,Anand2011,Passerini2008,Latora2008,Burda2010, Munoz2010,Coolen2011,Latora2011,Garnerone,Features, Severini2010a}. Methods for quantifying complexity are not only valuable from the theoretical point of view, but may lead to important operational interpretations. This new framework has the potential to  resolve  one of the outstanding problems in statistical mechanics of complex systems. In fact, it opens the way for a new information theory of complex network topologies which will provide an evaluation of the information encoded in complex networks.

 The entropy of network ensembles quantifies the number of graphs with given structural features such as degree distribution, degree correlations, community structure or spatial embedding  \cite{Bianconi2008a,Bianconi2008b,Bianconi2009,Anand2009,Anand2010,Annibale2009,Munoz2010,Coolen2011}. This quantity has been shown to be very useful for inference problems defined on networks  and it has been successfully applied to the problem of assessing the significance of features for network structure \cite{Features}. Other entropy measures of quantum mechanical nature have been  derived by mapping the network either to a density matrix or to a quantum state  \cite{BGS,BGS1,Passerini2008,Anand2011,Garnerone}. These entropies, defined on single networks, set a path for the application of tools of quantum information theory to describe the complexity of single networks and to introduce new kind of network parameters (for example, by considering the notion of correlations and  subsystems). Finally, entropy rate of random walks  \cite{Latora2008,Latora2011,Burda2010} are extensively studied on networks and they predict how evenly the random walk spreads  in the network. For many applications indeed one would wish to bias the random walk to construct maximally entropic random walks.

All these definitions of entropy of networks consider static networks, eventually explored by a dynamical process. In this paper we define   and evaluate the entropy rate of growing network models.
The literature in the field of growing network models generating scale-free networks is very large \cite{Albert2002, Dorogovtsev2003,Boccaletti2006,Dorogovtsev2010,Newman2010}.
By studying the entropy rate of these models we aim at quantifying the number of typical networks that are generated by these models and we compare this number with the number of networks that is possible to construct with the same degree distribution. In particular we focus on trees to allow for an analytic treatment of the problem. Trees are networks in which no cycle is allowed.
The maximal number of possible tree networks generated by a growing network model scales  like $N!$ where $N $ is the number of nodes (and links) in the network.
The minimal number of tree networks generated by a growing network model is one, corresponding to the formation of a star or of a linear chain.
The entropy rate of growing scale-free networks lies in between these two limiting values. Undestanding the value of the entropy of graphs is infomative because it describes the complexity of the growing network models. In fact the value of the entropy will quantify with a unique number the size  of the space of typical networks generated by the growing network model.
The smaller is the entropy rate of the networks the more complex the network structural properties implied by the growing model.
In particular it is essential to detremine the scaling with $N$ of the entropy rate, and in the case in which the entropy rate is not constant but depends on $N$ it is important to evaluate the subleading terms that encode for the topology of the networks also for other entropy measures \cite{Bianconi2008b,Bianconi2009,Garnerone}.

The main model of growing scale-free networks is the Barab\'asi-Albert (BA) model  \cite{BA} that generates scale-free networks with power-law exponent $ \gamma=3$. The BA networks are known to have weak degree correlations due to their causal structure, the growing network model with initial attractiveness \cite{attractiveness} of the nodes and the fitness model \cite{fitness} have more significant correlations. To quantify these correlations different measures have been introduced such as the average degree of the neighbor of the nodes or the degree correlation matrix.
Still we lack a way to quantify how much information is encoded in growing network models with respect  to the  networks constructed by the configuration model with the same degree distribution.

Here we propose to  quantify the number of typical tree graphs generated by the non-equilibrium growing network models \cite{BA, attractiveness, fitness, Bose, Redner1,Redner2, Aging} as a proxy of their complexity. This quantity can be used to measure the fraction of networks of given degree sequence that is generated by growing network models and to quantify in this way the complexity of growing network models.
Moreover growing network models as the  Bianconi and Barab\'asi  fitness model \cite{fitness,Bose} and the non-linear preferential attachment model of Krapivsky and Redner  \cite{Redner1, Redner2} or the growing network model with aging of the nodes  \cite{Aging} are known to undergo structural  phase transitions as a function of their   parameters.
Here we interestingly see that these phase transitions are characterized by a sharp drop of the entropy rate and strong finite size effects indicating that the network is reduced to a more ordered state.

The remainder of the paper is structured as follows. In Section II, we define the Gibbs entropy of networks with a given degree distribution. In Section III, we introduce the necessary material for studying the entropy rate of growing trees. Firstly, we recall the main growing network models. Then, we obtain min/max bounds to the entropy. In Section IV, we study growing trees with stationary degree distribution. In particular, we consider the BA model, initial attractiveness, the Bianconi-Barab\'asi fitness model, and networks with structural phase transitions. We draw some brief conclusions in Section V.
 \section{Gibbs entropy of networks with a given degree distribution}

The Gibbs entropy $ \Sigma[\{k_i\}]$  \cite{Bianconi2008a,Bianconi2008b,Bianconi2009,Anand2009,Anand2010} of a network ensemble with given (graphical) degree sequence $\{k_i\}$ \cite{DelGenio1,DelGenio2} is
 \begin{equation} \Sigma[\{k_i\}]=  \frac{1}{N} \log{ \cal N}[\{k_i\}]
 \end{equation} where ${ \cal N}[\{k_i\}]$ is the number of networks with the specified degree sequence and $N$ is the number of labeled nodes $i=1,2,\dots,N$. The Gibbs entropy depends on the number of links but also on the specif details of the degree sequence. In Table \ref{example} we give two illustrative examples for two  degree sequence compatible with   $5$ links but defining ensembles of networks with different entropy.
 
\begin{table}[ht]
\centering
	\begin{tabular} {|m{1.8cm}|m{4.5cm}|m{1.7cm}|}
		\hline
		\centering  Degree Sequence & \centering Networks &  \tc{Entropy}  \\
		\hline \hline
		\centering \includegraphics[scale=0.5]{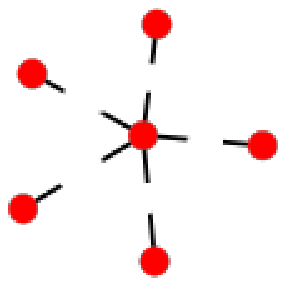} & \centering \includegraphics[scale=0.4]{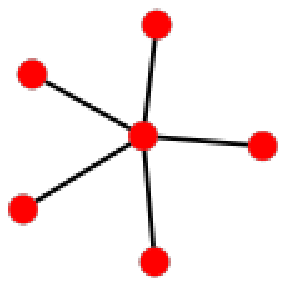} & \tc{$\Sigma[\{k_i\}] = 0$} \\					                           		\hline
		\centering \includegraphics[scale=0.5]{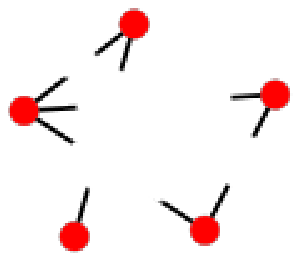} & \centering \includegraphics[scale=0.4]{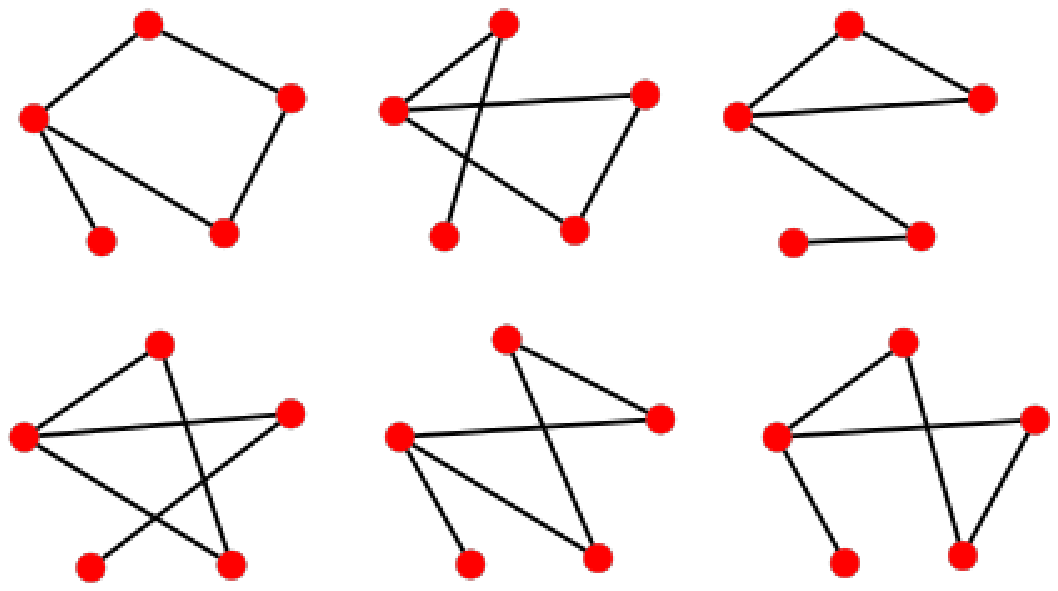} & \tc{$\Sigma[\{k_i\}] \neq 0$}\\
		\hline
	\end{tabular}
\caption{The configuration of networks with degree sequence \{1,1,1,1,5\} (on top, ${\cal N}[\{k_i\}]=1$) and \{1,2,2,2,3\} (on bottom, ${\cal N}[\{k_i\}]=6$).}
\label{example}
\end{table} 
 
It turns out that the ensemble of networks having a given degree distribution is a type of { \it microcanonical network ensemble}
satisfying a large number of hard constraints (the degree of each node is fixed). It is also possible to construct { \em canonical network ensembles} similar to what happens in classical statistical mechanics when one distinguishes the microcanonical and canonical ensembles according to the fact that  the energy is perfectly conserved or conserved in average.
A canonical network ensemble with given expected degree sequence is an ensemble of graphs in which the degree of each node is distributed as a Poisson variable with given expected degree $ \{ \overline{k}_i \}$.
The entropy of the canonical network ensemble is the logarithm of the typical number of networks in the ensemble. This entropy $S[\{\overline{k_i}\}]$  is given by
 \begin{equation}
S[\{\overline{k_i}\}]=- \frac{1}{N} \left[ \sum_{ij}p_{ij} \log p_{ij}- \sum_{ij}(1-p_{ij}) \log(1-p_{ij}) \right]
 \end{equation}
where $p_{ij}$ indicates the probability that a node $i$ is linked to a node $j$.
We can evaluate the entropy of a maximally random network ensemble with given expected degree distribuiton  $ \{ \overline{k}_i \}$  by maximizing the entropy  $S[\{\overline{k_i}\}]$ with respect to $p_{ij}$ under  the conditions
 \begin{equation}
 \overline{k}_i= \sum_j p_{ij}.
 \label{con}
 \end{equation}
In this way  we get for the marginal probabilities $p_{ij}$  \cite{Anand2009}
 \begin{equation}
p_{ij}= \frac{ \theta_i  \theta_j}{1+ \theta_i  \theta_j},
 \end{equation}
where $ \theta_i$ are related to the  lagrangian multipliers, or ``hidden variables'' fixed by the constraints given by Eqs. $( \ref{con})$.
In particular, for the uncorrelated network model in which  $\overline{k_i}< \sqrt{\avg{\overline{k}}N}$ and $p_{ij}=\frac{\overline{k_i} \overline{k_j}}{\avg{\overline{k}}N}$, the Shannon entropy network ensemble takes a direct form {\cite{Bianconi2009}}
 \begin{eqnarray}
\hspace*{-3mm} S[\{\overline{k_i}\}]&=& \frac{1}{2} \avg{k}[ \log( \avg{k}N)-1]- \frac{1}{N} \sum_i ( \ln \overline{k_i}-1)\overline{k_i}.
 \end{eqnarray}
The Gibbs entropy $ \Sigma$ of a microcanonical ensemble of networks with degree sequence $ \{k_i \}$ with $k_i= \overline{k_i}$ is given by  \cite{Anand2010}
 \begin{equation}
 \Sigma[\{k_i\}]=S[\{k_i\}]- \Omega[\{k_i\}]
 \end{equation}
where $ \Omega[\{k_i\}]$ is the entropy of large deviations of the canonical ensemble
 \begin{equation}
 \Omega[\{k_i\}]=- \frac{1}{N} \log[ \sum_{a_{ij}}p_{ij}^{a_{ij}}(1-p_{ij})^{1-a_{ij}} \prod_i  \delta( \sum_j a_{ij},k_i)].
 \end{equation}
where {$\{a_{ij}\} $ is the adjacency matrix  of the network. In particular the   matrix element  $a_{ij}$ of the adjacency matrix is    given by  $a_{ij}=1$ if a   link is present between node $i$ and node $j$ while  $a_{ij}=0$ otherwise.} By replica methods and the cavity method  \cite{Bianconi2008b,Anand2010} it is possible to derive the given expression for $ \Omega[\{k_i\}]$,
 \begin{equation}
 \Omega[\{k_i\}]=- \frac{1}{N} \sum_i  \log \pi_{k_i}(k_i),
 \end{equation}
where $ \pi_{r}(n)$ is the Poisson distribution with  $ \avg{n}=r$.
In particular, for the uncorrelated network model in which {$k_i< \sqrt{\avg{k}N}$ and $p_{ij}=\frac{k_i k_j}{\avg{k}N}$}, the Gibbs entropy network ensemble takes a direct form {\cite{Bianconi2009}}
 \begin{eqnarray}
 \Sigma[\{k_i\}]&=& \frac{1}{2} \avg{k}[ \log( \avg{k}N)-1]- \frac{1}{N} \sum_i ( \ln k_i-1)k_i+ \nonumber  \\
&&- \frac{1}{2N} \sum_i  \log (2 \pi k_i)- \frac{1}{4N} \left( \frac{ \avg{k^2}}{ \avg{k}} \right)^2.
 \end{eqnarray}
We might as well define the Gibbs entropy $\Sigma[\{N_k\}]$ of networks with given degree distribution $\{N_k\}$.
Since the number of graphs with given degree distribution ${\cal N}[\{N_k\}]$ is just given by
\begin{equation}
{\cal N}[\{N_k\}]={\cal N}[\{k_i\}] \frac{N!}{\prod_k N_k!}
\end{equation}
it follows that
\begin{equation}
\Sigma[\{N_k\}]=\Sigma[\{k_i\}]-\sum_k \frac{N_k}{N}\log\left(\frac{N_k}{N}\right).
\end{equation}
 \section{Entropy rate of growing trees}
Many networks are non static but they are growing by the addition of new nodes and links.
A major class of growing networks are growing trees in which at each time a new node and a new link is added to the network.
In the last ten years, many growing network models have been proposed. Special attention has been addressed to growing network models
generating scale-free networks. In fact these stylized models explain the basic mechanism according to which many growing natural networks develop the
universally found scale free degree distribution.
The fundamental model for scale-free growing network is the BA model  \cite{BA} which generates networks with degree distribution $P(k) \sim k^{- \gamma}$ and $ \gamma=3$.
This model is based of two ingredients: growth of the network and preferential attachment meaning that nodes with large degree are more likely to acquire new links.
In this paper we consider this model and other different significant variations to this model including different additional mechanisms as initial attractiveness of the nodes \cite{attractiveness}, fitness of the nodes\cite{fitness, Bose}, non-linear preferential attachment \cite{Redner1, Redner2} and aging of the nodes \cite{Aging}.
Some of these models as explained below undergo structural phase transitions to be studied by statistical mechanics methods.
 \subsection{Growing network models}
In the growing scale-free network model we start from two nodes linked together, at each time $t=1,2, \ldots$
 \begin{itemize}
 \item
we add a new node $i=t+2$;
 \item
we link the new node to a node $i_t$ of the network chosen with probability
 \begin{equation}
 \Pi(i_t)= \frac{A_{i_t}}{ \cal N},
 \end{equation}
where ${ \cal N}= \sum_{i=1}^{t+1} A_i$;
 \item
 the number of nodes in the network is $N=t+2$.
 \end{itemize}
As a function of the choice of $A_j$  different networks model are defined.
In particular we consider the following growing network models:
 \begin{itemize}
 \item
If we take $A_i= \delta_{i,1}$, we get  a star graph;
 \item
If we take $A_i= \delta_{i,t+1}$, we get the  linear chain;
 \item
If we take $A_i=1$, we get a maximally random connected and growing tree;
 \item
If we take $A_i=k_i$, where $k_i$ is the degree of the node $i$, we get the BA model  \cite{BA};
 \item
If we take $A_i=k_i-1+a$, with $a<1$, we get a generalized BA model with initial attractiveness of the nodes  \cite{attractiveness};
 \item
If we assign to each node a fitness value $ \eta_i$ from a distribution $ \rho( \eta)=1$ and $ \eta  \in (0,1)$ and we take $A_i= \eta_ik_i$, we get the  Bianconi-Barab\'asi fitness model  \cite{fitness}.
 \item
If we take $A_i=k_i^{ \gamma^{ \prime}}$, we get the non-linear preferential attachment model of Krapivsky-Redner  \cite{Redner1,Redner2}. This network model undegoes a gelation phenomenon for $ \gamma^{ \prime}>1$.  Namely, there is an emergence of a single dominant node  linked to almost every
other node. For $ \gamma^{ \prime}>2 $,
 there is a finite probability that the dominating node is the first node of the network.
  \item
If we assign to each node a fitness value $ \eta_i=e^{- \beta  \epsilon_i}$, with $ \epsilon_i$ drawn from a distribution $g( \epsilon)\propto \epsilon^{\kappa}$ and $\kappa>0$, and we take $A_i= \eta_ik_i$, we get as a function of $ \beta$ the so called "Bose-Einstein condensation in complex networks" of Bianconi and Barab\'asi  \cite{Bose}. When this happens, for  $ \beta> \beta_c$  one node with high fitness is connected to a finite fraction of other nodes in the network.
  \item
If we take $A_i=(t-t_i)^{- \alpha}k_i$ where $t_i$ indicates the time at which the node $i$ has joined the network,  we get the preferential attachment model with aging of the sites of  Dorogovtsev-Mendes  \cite{Aging}. In this growing network the power-law exponent $\gamma $ of the degree distribution is diverging as $ \gamma \simeq  \frac{1}{c_1} \frac{1}{1- \alpha}$ when $ \alpha \to 1^-$. For $ \alpha>1$ the network is exponential and becomes more and more similar to a linear chain.

 \end{itemize}
 \subsection{Entropy rate}
The growing connected trees are fully determined by the sequence of symbol ${ \cal X}=(i_1,i_2, \ldots, i_N)$ where $i_{t}$ is the node linked at time $t$ to the node $i=t+2$.
In order to evaluate the entropy rate of growing networks it is sufficient to determine the entropy rate of the sequence $(i_1,i_2 \ldots, i_t)$:
 \begin{equation}
{h(t,{ \cal X})}=-{ \sum_{i_t} P(i_t|i_1,i_2, \ldots i_{t-1}) \log P(i_t|i_1,i_2, \ldots, i_{t-1})},
 \end{equation}
where $P(i_t|i_1,i_2, \ldots i_{t-1})$ is the conditional probability that the node $i_t$ is chosen at time $t$ given the history of the process.
The entropy of the process evaluating how many networks are typically constructed by the growing network process is $S( \{{ \cal X} \})$
 \begin{equation}
{S}( \{{ \cal X} \})=- \sum_{ \{i_1,i_2, \ldots , i_t  \}}P(i_1,i_2, \ldots i_t) \log P(i_1,i_2, \ldots , i_t).
 \end{equation}

 \subsection{Maximal and minimal bound to the entropy rate of growing network trees}
It is instructive to study the limits of the entropy rate of connected growing trees.
The minimal entropy rate is given by the entropy rate of the star or of the linear chain. Indeed by taking $A_i= \delta_{1,i}$ we have that the entropy rate is zero. Indeed the growing network model becomes deterministic and it gives rise to a unique star network with the center on the node $i=1$.
The entropy rate  of a linear chain $A_i= \delta_{i,t+1}$ is also zero by a similar argument and the model generates a unique linear chain network structure.
On the other hand the maximal entropy rate is given by the maximally random growing connected trees that is generated by taking $A_i=1$ and $ \Pi_i=1/(t+1)$.
For this process the entropy rate is given by
 \begin{equation}
h(t,{ \cal X})= \log (t+1)
 \end{equation}

Therefore this entropy rate increases logarithmically with time
and the probability of each tree with $N=t+2$ nodes is given by
 \begin{equation}
P(N)= \frac{1}{(N-1)!}
 \end{equation}
Therefore  $S( \{{ \cal X} \})= \log[(N-1)!]$.
This is the maximal entropy of a growing connected tree.
 \section{Growing trees with stationary degree distribution}

For growing network models with stationary degree distributions there are simple relations between $h(t,{ \cal X})$ and $S( \{ \cal X \})$.
Indeed let us define the  entropy  rate
 \begin{equation}
{H}({ \cal X})= \lim_{N \to  \infty} \frac{1}{N}[S( \{{ \cal X} \})- \log[(N-1)!].
 \end{equation}
For a growing tree network with stationary degree distribution, by the recursive application of chain rule $P(i_1,i_2,\ldots, i_t)=P(i_t|i_1,i_2,\ldots i_{t-1})P(i_1,i_2,\ldots, i_{t-1})$  we  can easily get
 \begin{equation}
{H}({ \cal X})= \lim_{N \to  \infty} \frac{1}{N}[ \sum_{n=1}^{N-2} h(n,{ \cal X})- \log((N-1)!)].
 \end{equation}
If the entropy rate of growing networks $H$ is a constant, it means that the number of graphs generated by the growing network model has a dominating term which goes like $N!$ and a subleading term that is exponential with the number of nodes $N$. On the contrary if $H=- \infty$ it means that the number of networks generated by the growing network model increases with the number of nodes in the network $N$ at most exponentially.
Usually the  typical number of labeled networks generated by growing network models with convergent  degree distribution is less than the number of networks with the same  degree distribution.
In order to evaluate the ratio between these two cardinalities,  we introduce here  the difference $ \Delta$ between the Gibbs entropy $ \Sigma[\{N_k\}]$ of the network with the same degree distribution and the entropy of the networks generated by the growing network model.
Therefore $ \Delta$ is
 \begin{equation}
 \Delta= \lim_{N \to  \infty} \left \{ \Sigma[\{N_k\}]- \frac{1}{N} \log[(N-1)!]-{{H({ \cal X})}} \right \}.
 \end{equation}
The larger the value of $ \Delta$ is, the smaller the fraction of networks generated by the growing model compared with the networks generated by the configuration model. This implies that the larger is $\Delta$ the more complex the networks generated by the growing models are. In fact these networks need  the dynamics of the networks implicitly force the networks to satisfy more stringent set of structural conditions  beyond the degree distribution.

 \subsection{The entropy rate of the BA model}
We consider the BA model,  we take $A_i=k_i$ and $ \Pi_i= \frac{k_i}{2(N-1)}$ therefore
 \begin{equation}
P(i_t|i_1,i_2, \ldots, i_{t-1})= \frac{k_t}{2(N-1)}.
 \end{equation}
The BA model, asymptotically in time has a degree distribution that converges to the value $N_k$ given by
 \begin{equation}
N_k= \frac{{4}N}{k(k+1)(k+2)},
\label{asym}
 \end{equation}
Therefore, asymptotically in time the entropy rate of the BA model is
 \begin{equation}
{h(t=N-2,{ \cal X})} \to- \sum_{k=1}^{ \infty} N_k  \frac{k}{2(N-1)} \log \left( \frac{k}{2(N-1)} \right).
 \end{equation}
Hence, the entropy rate $h({ \cal X})$ increases in time as the logarithm of the number of nodes in the network, but it has a subleading term which is constant in time and depends on the degree sequence, \emph{i.e.}
 \begin{eqnarray}
{h(t=N-2,{ \cal X})}& \to & \log(N-1)+ \log(2)  \nonumber  \\
&&- \sum_{k=1}^{N-1}  \log(k) \frac{2}{(k+1)(k+2)} \nonumber  \\
& \to & \log(N-1)-0.51(0)
 \end{eqnarray}
The entropy rate ${H({ \cal X})}$ in the limit $N \to  \infty$ is therefore given by
 \begin{equation}
{H}({ \cal X})\simeq-0.51\ldots.
 \end{equation}

We note here that the degree distribution $N_k$ is known to have interesting finite size effects\cite{Dorogovtsev2003,Godreche}, in addition to the asymptotic scaling Eq.$ (\ref{asym})$. Here we checked that the value of the entropy rate is not modified by these corrections up to the significant digit we have considered.
Finally, in order to compare the number of networks generated by the BA model with the network that we can construct with the same degree sequence, we  evaluate the value of  $ \Delta$ in the thermodynamic limit. This can be written as
 \begin{eqnarray}
 \Delta&=& \lim_{N \to  \infty} \left \{ \frac{1}{N}  \sum_{t=1}^{N-1} \log(N/t) \right. \nonumber \\
&&\left.-\frac{1}{2N} \sum_k N_k [k\log(k)+\log(2\pi k)] \right. \nonumber  \\
&&\left.-\sum_k \frac{N_k}{N}\log\left(\frac{N_k}{N}\right)+1\right\} \nonumber \\
&& \simeq 0.9(1)
 \end{eqnarray}

 \subsection{The entropy rate of the growing network model with initial attractiveness}
\begin{figure}
 \includegraphics[width=0.6 \columnwidth, height=40mm]{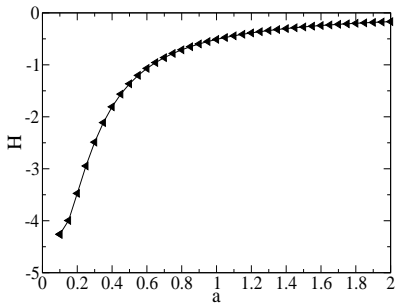}
 \caption{The entropy rate $H$ calculated for the growing network model with initial attractiveness \cite{attractiveness} as a function of $a$ and evaluated by Eq. $(\ref{Ha})$ using a maximal degree equal to $K=10^7$.}
 \label{fig1}
 \end{figure}

\begin{figure}
 \includegraphics[width=0.6 \columnwidth, height=40mm]{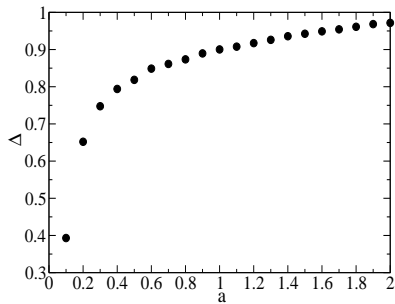}
 \caption{The value of $\Delta $ calculated for the growing network model with initial attractiveness \cite{attractiveness} as a function of $a$ evaluated for networks of $N=50000$ nodes and over $20$ realizations of the process.}
 \label{fig_delta}
 \end{figure}

If we take $A_i=k_i-1+a$, the network generated is scale free with power-law exponent $ \gamma=2+a$ \cite{attractiveness}.
The probability to choose the node $i_t$ given the history of the process is therefore given by

 \begin{equation}
P(i_t|i_1,i_2, \ldots, i_{t-1})= \frac{A_i}{ \cal N}= \frac{k_i-1+a}{(a+1)(N-1)}.
 \end{equation}
Asymptotically in time the degree distribution for trees converges to the value  \cite{Dorogovtsev2003, attractiveness}
 \begin{equation}
N_k=N(1+a) \frac{ \Gamma(1+2a) \Gamma(k+a-1)}{ \Gamma(a) \Gamma(k+1+2a)}.
 \end{equation}
From this, the  entropy rate $h(t,{ \cal X})$ is asymptotically
 \begin{eqnarray}
{h(t=N-2,{ \cal X})} \to \nonumber \\
&&\hspace*{-35mm}- \sum_{k=1}^{ \infty} N_k  \frac{k-1+a}{(a+1)(N-1)} \log \left[ \frac{k-1+a}{(a+1)(N-1)} \right],
 \end{eqnarray}
which can be simplified as
 \begin{eqnarray}
{h(t=N-2,{ \cal X})}& \to & \log(N-1)+ \log(a+1)  \nonumber \\
&&\hspace*{-12mm}- \sum_{k=1}^{N-1}  \log \left[(k-1+a) \frac{ \Gamma(1+2a) \Gamma(k+a)}{ \Gamma(a) \Gamma(k+1+2a)} \right]. \nonumber  \\
 \end{eqnarray}
In the limit $N  \to  \infty$, the entropy rate ${H({ \cal X})}$ is
 \begin{eqnarray}
 {H}({ \cal X})=  \log(a+1) \nonumber \\
 &&\hspace*{-25mm}- \sum_{k=1}^{N-1} \log \left[(k-1+a) \frac{ \Gamma(1+2a) \Gamma(k+a)}{ \Gamma(a) \Gamma(k+1+2a)} \right].
 \label{Ha}
 \end{eqnarray}
When $a  \to 1$, the solution reduces to the solution of the  BA model. In Fig.  \ref{fig1} we plot   the value of $H={H}({ \cal X})$ versus  $a$ calculated by Eq. (\ref{Ha}) using an upper cutoff for the degree $k_i<K \forall i=1,\ldots N$. As the parameter $a \to 0$ the entropy rate decreases indicating that the network model generates an exponentially smaller number of networks. Also  the Gibbs entropy of  scale free networks decreases as long as the power-law exponent converges toward 2, \emph{i.e.} in the limit $ \gamma \to 2$.  In order to evaluate the change in the ratio of networks generated by the growing network model to the number of possible networks
with the same degree distribution, in Fig. $\ref{fig_delta}$ we plotted $\Delta$ as a function of $a$. As $a\rightarrow 0$ and $\gamma\to 2$ the number of networks generated by the growing network models are a smaller function of the total number of networks that is possible to build with the same degree distribution. This is an indication and quantification of the importance of correlations generated by the growing network model with given initial attractiveness $a$.
 \subsection{The entropy rate of the Bianconi-Barab\'asi fitness model}
If the kernel $A_i$ is heterogeneous and specifically given by $A_i= \eta_i k_i$, the model is called Bianconi-Barab\'asi fitness model  \cite{fitness}. The probability to choose the node $i_t$ given the history of the process is therefore given by
 \begin{equation}
P(i_t|i_1,i_2, \ldots, i_{t-1})= \frac{A_i}{ \cal N}= \frac{ \eta_i k_i}{ \mu (N-1)},
 \end{equation}
where $ \mu (N-1)= \sum_{i=1}^{N-1} \eta_i k_i$.
The degree distribution $N_k( \eta)$ for nodes of fitness $ \eta$, asymptotically in time  converges to  \cite{Newman2010}
 \begin{equation}
N_k( \eta)= \frac{N \mu \rho( \eta)}{ \eta} \frac{ \Gamma(k) \Gamma(1+ \mu/ \eta)}{ \Gamma(k+1+ \mu/ \eta)},
 \end{equation}
where $ \rho( \eta)$ is the distribution of $ \eta$.
Given the analytic solution of the model  \cite{fitness, Newman2010},  $ \mu$ is determined by the self-consistent relation
 \begin{equation}
 \int_0^{ \eta_0} \rho( \eta)( \mu/ \eta-1)^{-1}d \eta=1
 \end{equation}
We consider here the case of uniform distribution of the fitness, \emph{i.e.}  $ \rho( \eta)=1$ with  $ \eta  \in (0,1)$. Therefore the entropy rate is given by
 \begin{eqnarray}
h(t=N-2,{ \cal X}) \to \nonumber \\
&&\hspace*{-30mm}- \sum_{k=1}^{N-1} \int_0^{1}N_k( \eta) \frac{ \eta k}{ \mu (N-1)} \log \left[ \frac{ \eta k}{ \mu (N-1)} \right],
 \end{eqnarray}
which gives
 \begin{equation}
{H}({ \cal X}) = -1.59  \dots
 \end{equation}

\subsection{ Entropy rate for growing  network models  with structural  phase transitions}
We have measured the entropy rate for three growing network models showing a phase transition:
 \begin{itemize}
 \item The Krapivsky-Redner model  \cite{Redner1, Redner2} with
 \begin{equation}
h(t,{ \cal X}) = -\sum_{i=1}^t\frac{k_i^{\gamma'}}{\cal N} \log \left( \frac{k_i^{\gamma'}}{\cal N} \right)
\end{equation}
and ${\cal N}=\sum_{i=1}^{t}k_i^{\gamma'}$
 \item
The Bianconi-Barab\'asi   model showing  a Bose-Einstein condensation in complex networks  \cite{Bose}
with
\begin{equation}
h(t,{ \cal X}) = -\sum_{i_t}\frac{e^{-\beta \epsilon_i} k_i}{\cal N} \log \left( \frac{e^{-\beta \epsilon_i} k_i}{\cal N} \right)
\end{equation}
and ${\cal N}=\sum_{i=1}^{t}e^{-\beta \epsilon_i}k_i$.
 \item The Dorogovtsev-Mendes  model with aging of the nodes  \cite{Aging}
 {with}
\begin{equation}
h(t,{ \cal X}) = -\sum_{i_t}\frac{\tau_i^{-\alpha} k_i}{\cal N} \log \left( \frac{\tau_i^{-\alpha} k_i}{\cal N} \right)
\end{equation}
where $\tau_i = t - t_i$ is the age of node $i$ and ${\cal N}=\sum_{i=1}^{t}\tau_i^{-\alpha}k_i$.
\end{itemize}
In Fig. \ref{fig2}  the entropy rate $H(\cal X)$ is calculated {by numerical simulations using
 \begin{equation}
{H}({ \cal X})=  \frac{1}{N}[ \sum_{n=1}^{N-2} h(n,{ \cal X})- \log((N-1)!)].
 \end{equation} for a network of sufficiently large size $N$ for the three models as a function of  { the parameters $\gamma$, $\beta$ and $\alpha$ respectively.}
We show that at the transition point the scaling of ${H}$ evaluated for a network of  size $N$ changes from constant  to an $N$ dependent behavior.
In particular we checked that in the three cases $H\propto \log(N)$ indicating that as the network grows the typical number of networks that are generated scales only exponentially  with $N$ (and not like $N!$).

 This behavior signifies a disordered-ordered phase-transition in the topology of the network. In the Bose-Einstein condensation network model and in the Krapivsky-Redner model, below the phase transition, the network is dominated by a hub node that grabs a finite fraction of the nodes. In the aging model, below the phase transition, the network develops a structure more similar to a linear chain.
\begin{figure*}
 \includegraphics[width=0.8 \columnwidth, height=100mm]{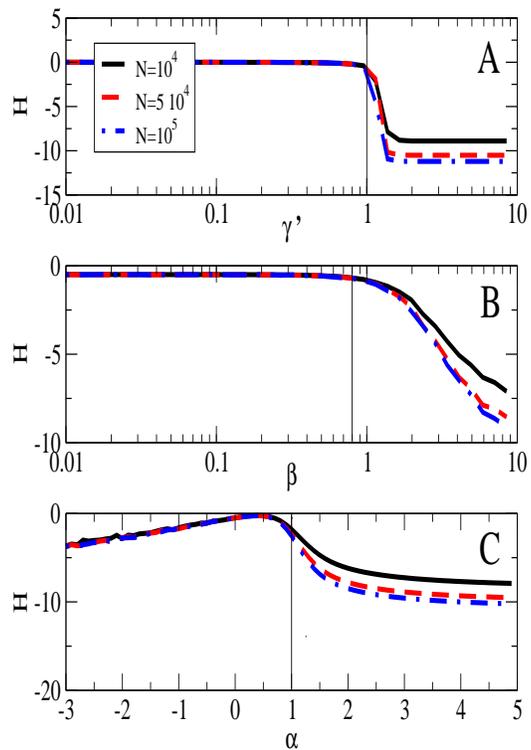}
 \caption{(Color online) The entropy rate $H$ is evaluated  for the Kapivsky-Redner model \cite{Redner1, Redner2} (panel A),  for the  "Bose-Einstein condesation in complex networks" of Bianconi-Barab\'asi  with  $g(\epsilon)=2\epsilon$, and $\epsilon\in(0,1)$, $(\kappa=1)$ \cite{Bose} (panel B) and for the aging model \cite{Aging} of Dorogovtsev-Mendes (panel C). The data are averaged over $N_{run}$ different realizations of the network . We took $N_{run}=100$ for simulations with $N=10^4$ and $N_{run}=30$ otherwise. Above the structural phase transition indicated with the solid line, the entropy rate $H$ strongly depends on $N$.}
 \label{fig2}
 \end{figure*}
 
 \section{Conclusion}
In conclusion, we have studied growing network models and their entropy rate.
We have seen that the entropy rate of growing simple trees have maximal and minimal bound and we have studied the entropy rate of scale-free tree networks.
This entropy rate allows us to calculate the number of typical graphs generated by growing scale-free network models and to quantify their complexity by comparing this number to the total number of graphs with the same degree distribution.
Although we have focused on trees the definition of entropy rate can be easily extended to growing network models with cycles. However the probabilities of adding two or more links at a given time should explicetly account for the fact that the new  links must be distinct, fact which induces a small correction to the simple preferential attachment.
We have analyzed  a variety of growing network models and  we have studied non-equilibrium growing network models showing structural  phase transitions. By numerical investigations, we  have shown that when a growing network model has a phase transition, the entropy rate changes its scaling with the system size indicating the disorder-to-order transition.
In the future, we believe that an integrated view of information theory of complex networks will provide a framework to extend quantitative measures of complexity to a large variety of network structures, models and dynamics. The present work is a step in this direction.

 \end{document}